\documentclass[fleqn,usenatbib]{mnras}

\usepackage{newtxtext,newtxmath}
\usepackage{graphicx}

\usepackage[T1]{fontenc}

\DeclareRobustCommand{\VAN}[3]{#2}
\let\VANthebibliography\thebibliography
\def\thebibliography{\DeclareRobustCommand{\VAN}[3]{##3}\VANthebibliography}

\usepackage{graphicx}	
\usepackage{amsmath}	

\title[Haralick features]{Rapid sorting of radio galaxy morphology using Haralick features}

\author[Ntwaetsile \& Geach]{
Kushatha Ntwaetsile\thanks{E-mail: k.ntwaetsile@herts.ac.uk} and
James E. Geach
\\
Centre for Astrophysics Research, Department of Physics, Astronomy \& Mathematics, University of Hertfordshire, College Lane, Hatfield, AL10 9AB, UK\\
}

\pubyear{2020}

\begin{document}
\label{firstpage}
\pagerange{\pageref{firstpage}--\pageref{lastpage}}
\maketitle

\begin{abstract}
We demonstrate the use of Haralick features for the automated classification of radio galaxies. The set of thirteen Haralick features represent an extremely compact non-parametric representation of image texture, and are calculated directly from imagery using the Grey Level Co-occurrence Matrix (GLCM). The GLCM is an encoding of the relationship between the intensity of neighbouring pixels in an image. Using 10,000 sources detected in the first data release of the LOFAR Two-metre Sky Survey (LoTSS), we demonstrate that Haralick features are highly efficient, rotationally invariant descriptors of radio galaxy morphology. After calculating Haralick features for LoTSS sources, we employ the fast density-based hierarchical clustering algorithm HDBSCAN to group radio sources into a sequence of morphological classes, illustrating a simple methodology to classify and label new, unseen galaxies in large samples. By adopting a `soft' clustering approach, we can assign each galaxy a probability of belonging to a given cluster, allowing for more flexibility in the selection of galaxies according to combinations of morphological characteristics and for easily identifying outliers: those objects with a low probability of belonging to any cluster in the Haralick space. Although our demonstration focuses on radio galaxies, Haralick features can be calculated for any image, making this approach also relevant to large optical imaging galaxy surveys.  
\end{abstract}

\section{Introduction}

We are entering the era of `big' observational astronomy, with surveys such as the Dark Energy Survey \citep[DES][]{proposal54}, Vera Rubin Observatory Legacy Survey of Space and Time \citep[LSST][]{proposal95}, {\it Euclid} \citep[]{proposal90}, Australian Square Kilometer Array Pathfinder (ASKAP) \citep[]{proposal93}. In these large surveys, simply curating the data volume becomes a significant challenge and interesting scientific question in its own right. For example, in the radio, the Square Kilometer Array \citep[SKA][]{proposal53} will be the largest big data project in the world, producing 1000 petabytes of data each day and about 5 {\it zettabytes} of data per year \citep{paper2}. Like all the large surveys of present and future, within the vast datasets lie new parameter spaces and new discoveries to be made. 

Machine learning is becoming an important statistical tool for astronomers seeking to efficiently analyse and derive meaning from massive data sets, with `unsupervised' methods in particular showing promise in various applications, particularly in automatic classification, including the separation of starbursts from active galactic nuclei \citep[]{proposal58}, optical morphology \citep[]{proposal49,proposal47}, variable stars \citep[]{paper12,paper11} and light curves \citep[]{paper13}. Other key applications are in parameter estimation (e.g.\ redshifts, \cite[][]{proposal58}), anomaly and outlier detection \citep[]{paper10,paper8}. Although not without limitations, unsupervised techniques are attractive for astronomical data exploration because they do not require labelled training sets, unlike, for example, neural network-based classification approaches \citep[e.g.][]{proposal60, proposal61}. Instead, unsupervised techniques generally rely on the data itself to identify patterns (e.g.\ clusters of self-similar objects) in some pre-defined `feature space' that describes the {\it n}-dimensional volume where the data are defined. At a simple level, this could just represent a vector of observations (e.g.\ photometry across a set of filters), but often it is desirable to project the data into another space, and that choice is a human one. 

In astronomy the two main types of data usually encountered are catalogues \citep[e.g.][]{proposal62}, and imaging. Working with catalogues is efficient, since they are a significantly reduced representation of the original data; for example in many astronomical images the majority of the pixels do not contain useful signal (at least not at first glance). On the other hand, working with the imaging data directly is preferred in some cases, since certain information is not easily captured, or completely missing from catalogues. An example of this is in morphological classification of galaxies. Although there are various simple parametric measures of morphology based on the distribution of pixel brightness (e.g. Gini, $M_{20}$ \citep[]{proposal63} ; CAS \citep[]{proposal64}), unsupervised machine learning offers opportunities to extract more fine detail -- and perform automatic classification -- directly from imaging data.  

For example, \cite{proposal65} use Principal Component Analysis (PCA) to project {\it Hubble Space Telescope} thumbnail images of high redshift galaxies into a 12-dimensional space where galaxies are represented as weighted combinations of `eigengalaxies'. The authors show how the set of weights for each galaxy can be clustered to produce meaningful morphological classifications that ultimately required no human intervention to partition. In another example, \cite{proposal49} and \cite{proposal47} present an approach that uses a hierarchical clustering technique to group together small image patches into self-similar clusters, representing each image patch as a vector containing the Fourier transform of each patch to represent the spatial frequency distribution of pixel brightness in several optical bands. The clustering resulted in a set of visually pure morphological classes, with properties that can be linked directly with crowd sourced (i.e.\ human) classifications from the Galaxy Zoo project \citep[]{proposal66}, illustrating the power and potential of unsupervised techniques in producing interpretable morphological classifications directly from imaging data. This is important because even with mega crowd sourcing, it is unlikely that a large cohort of citizen scientists \citep[]{article67} will be able to classify the billions of objects to be detected by the likes of LSST, {\it Euclid} and SKA. Carbon footprint aside, machines are inexhaustible and consistent. 

Classification of galaxy morphology has been performed over the years using a wide variety of techniques of increasing sophistication. Traditionally visual inspection of galaxies in the optical was carried out using Hubble's classification scheme \citep[]{proposal89,proposal74}. For clusters of galaxies, the Bautz--Morgan (BM) system was developed \citep[]{proposal68} to classify the morphology of clusters according to the difference in apparent magnitude between the brightest and next brightest members of the cluster. In the radio, \cite{proposal78} introduced the famous Fanaroff-Riley classification scheme that divides radio galaxies into two classes based on the relative brightness distribution of within the radio lobes and jets. Fundamentallty, these early classification techniques across the electromagnetic spectrum were based on the relative distribution of pixel (or photographic plate exposure) brightness, with human beings ultimately performing the classification.

As astronomy became increasingly data driven over the twentieth century, automated morphological classifications using machine learning were introduced. Artificial neural networks were trained to classify galaxies based on their morphology from the early 1990s \citep[]{proposal73,proposal72,proposal75}. Since then an increasing number of studies have applied machine learning for the classification of radio galaxy morphologies including the use of Self Organised Maps \citep[]{proposal84, proposal85,proposal94} and Convolutional Neural Networks \citep[]{proposal76,proposal77,proposal82}. More recently, `capsule networks' have also been used for the morphological classification of radio galaxies \citep[]{proposal79, proposal86}. In this work we contribute to the effort of seeking more {\it efficient} means of automatically classifiying and mining large imaging data sets in the radio using `Haralick' features \citep{proposal28}, which provide a compact representation of image texture. Although a well-established tool in computer vision, Haralick features have not yet been widely exploited in radio astronomy. \citep[although see][]{proposal27}.

The paper is organised as follows: an introduction to Haralick features is presented in Section 2, in Section 3 we describe the data set we use as a demonstration of the technique from the LOFAR Two-meter Sky Survey, and the Hierarchical Density Based Spatial Clustering of Applications with Noise (HDBSCAN) algorithm we adopt to perform classification. We present the results in Section 4 and make our conclusions in Section 5.

\section{Haralick features}

Haralick features were introduced by \cite{proposal28} as a statistical method of examining image texture. They consider the relationship between the intensity of neighbouring pixels and are useful in classifying similar regions in an image given the local spatial distribution of pixel intensities\citep{proposal30}. Haralick features are based on the so-called Grey Level Co-occurrence Matrix (GLCM). To compute the Haralick features, one must first compute the GLCM, which can then be used to evaluate the Haralick coefficients \citep[]{proposal29}. 

The GLCM $p$ is a square matrix where each dimension corresponds to the number of greyscale intensity values in the image. For an 8-bit depth, this would result in a 256$^2$ matrix, so the bit depth is normally reduced to 3, such that $p$ is an $8\times8$ matrix. Each element $p(i,j)$ is calculated as the frequency that a pixel with intensity $i$ has a pixel with intensity $j$ at a given spatial offset. This offset is normally the pixel in the same row but adjacent column, but other offsets can be used. Therefore $p$ encodes the relative spatial distribution of greyscale levels in an image. Mathematically, for an image $I$ of size $M\times N$ pixels:  

\begin{equation*} \label{} p(i, j)=\sum_{r=1}^{M}\sum_{c=1}^{N}\begin{cases} 1~~\mathrm{if}~I(r,c)=i\ \mathrm{and}~I(r+\Delta x,c+\Delta y)=j\\ 0~~\mathrm{otherwise} \end{cases} \tag{1} \end{equation*}
where $\Delta x$ and $\Delta y$ are the desired offset. Typically we have ($\Delta x=1, \Delta y=0$), however calculating $p$ for the different directional offsets and averaging, rotational invariance can be encoded. Finally, $p$ is normalised by the total number of comparisons made such that the GLCM represents a probability distribution.

With $p$ defined, we can define 14 Haralick features that describe the textural properties of the image, or rather, an image section \citep{proposal28,proposal40}. The features are defined as follows:

\medskip

\noindent {\it Angular Second Moment}

      \begin{equation}\label{}
     f_1=\sum_{i}\sum_{j}p(i, j)^2
     \end{equation}
     
\smallskip

\noindent {\it Contrast}

    \begin{equation} \label{}
          f_2=\sum_{i}\sum_{j}(i-j)^{2}p(i, j) 
    \end{equation}

\smallskip

\noindent {\it Entropy}

    \begin{equation}\label{}
    f_3=-\sum_{i} \sum_{j}p(i, j)\log(p(i, j))
    \end{equation}
    
\smallskip

\noindent {\it Correlation}

    \begin{equation}\label{}
    f_4= \frac{1}{\sigma_{x}\sigma_{y}}\sum\limits_{i}\sum\limits_{j}(ij)p(i, j)-\mu_{x}\mu_{y}
    \end{equation}
where $\mu_{x,y}$ and $\sigma_{x,y}$ are the means and standard deviations of the corresponding row or column in $p$:

\begin{equation}
    \mu_x = \sum_{ij} ip(i,j)
\end{equation}

\begin{equation}
    \mu_y = \sum_{ij} jp(i,j)
\end{equation}

\begin{equation}
    \sigma_x = \sqrt{\sum_{ij} (i-\mu_x)^2p(i,j)}
\end{equation}

\begin{equation}
    \sigma_y = \sqrt{\sum_{ij} (j-\mu_y)^2p(i,j)}
\end{equation}

    \smallskip

\noindent {\it Variance}
     
     \begin{equation}\label{}
      f_5=\sum_{i}\sum_{j}(j-\mu)^{2}p(i, j)
      \end{equation}
      
\smallskip

\noindent {\it Homogeneity (a.k.a. Inverse Difference Moment)}

      \begin{equation}\label{}
      f_6=\sum_{i}\sum_{j}{p(i, j) \over 1+(i-j)^{2}}
      \end{equation}

\smallskip

\noindent {\it Sum Average}
      
      \begin{equation}\label{}
      f_7=\sum\limits_{i=2}^{2{N}}i p_{x+y}(i)
    \end{equation}
where $N$ is the number of grey levels and $p_{x+y}(i)$ corresponds to

\begin{equation}
    p_{x+y}(i) = \sum_{x+y=i}p(i,j)
\end{equation}

\smallskip

\noindent {\it Sum Entropy}

       \begin{equation}\label{}
      f_8=\sum\limits_{i=2}^{2{N}}p_{x+y}(i)\log(p_{x+y}(i))
    \end{equation}
    
\smallskip

\noindent {\it Sum Variance}

      \begin{equation}\label{}
      f_9=\sum\limits_{i=2}^{2{N}}(i-f_8)^{2}p_{x+y}(i)
    \end{equation}

\smallskip

\noindent {\it Difference Entropy}
    
       \begin{equation}\label{}
      f_{10}=-\sum\limits_{i=0}^{N-1}p_{x-y}(i)\log(p_{x-y}(i))
    \end{equation}
where
    \begin{equation}
    p_{x-y}(i) = \sum_{x-y=i}p(i,j)
\end{equation}

\smallskip

\noindent {\it Difference Variance}
    \begin{equation}\label{}
      f_{11} = \sum_{i=0}^{N-1} i^2p_{x-y}(i)
    \end{equation}

\smallskip
  
\noindent {\it Information Measure of Correlation 1}
      \begin{equation}\label{}
      f_{12}={\text{HXY}-\text{HXY1}\over \text{max\{HX, HY\}}} 
    \end{equation}

\smallskip
  
\noindent {\it Information Measure of Correlation 2}

        \begin{equation}\label{}
      f_{13}=\sqrt{1-\exp(-2(\text{HXY2}-\text{HXY})}
    \end{equation}
where HX and HY are entropies of $p_x$ and $p_y$:

    \begin{equation}
    \text{HXY}=-\sum\limits_{i}\sum\limits_{j}p(i, j)\log(p(i, j))
    \end{equation}
    \begin{equation}
    \text{HXY1}=-\sum\limits_{i}\sum\limits_{j}p(i, j)\log(p_{x}(i)p_{y}(j))
    \end{equation}
    \begin{equation}
    \text{HXY2}=-\sum\limits_{i}\sum\limits_{j}p_{x}(i)p_{y}(j)\log(p_{x}(i)p_{y}(j))
    \end{equation}

\smallskip
   
\noindent {\it Maximal Correlation Efficient}

\smallskip

\noindent With the definition 
    \begin{equation}\label{}
        Q(i,j)=\sum_k \frac{p(i,k)p(j,k)}{p_x(i)p_y(k)}
    \end{equation}
The maximal correlation coefficient $f_{14}$ is defined as the square root of the second largest eigenvalue of $Q$. This is generally considered to be numerically unstable and we only consider the first 13 Haralick features.

\begin{figure}
  \includegraphics[width=0.5\textwidth]{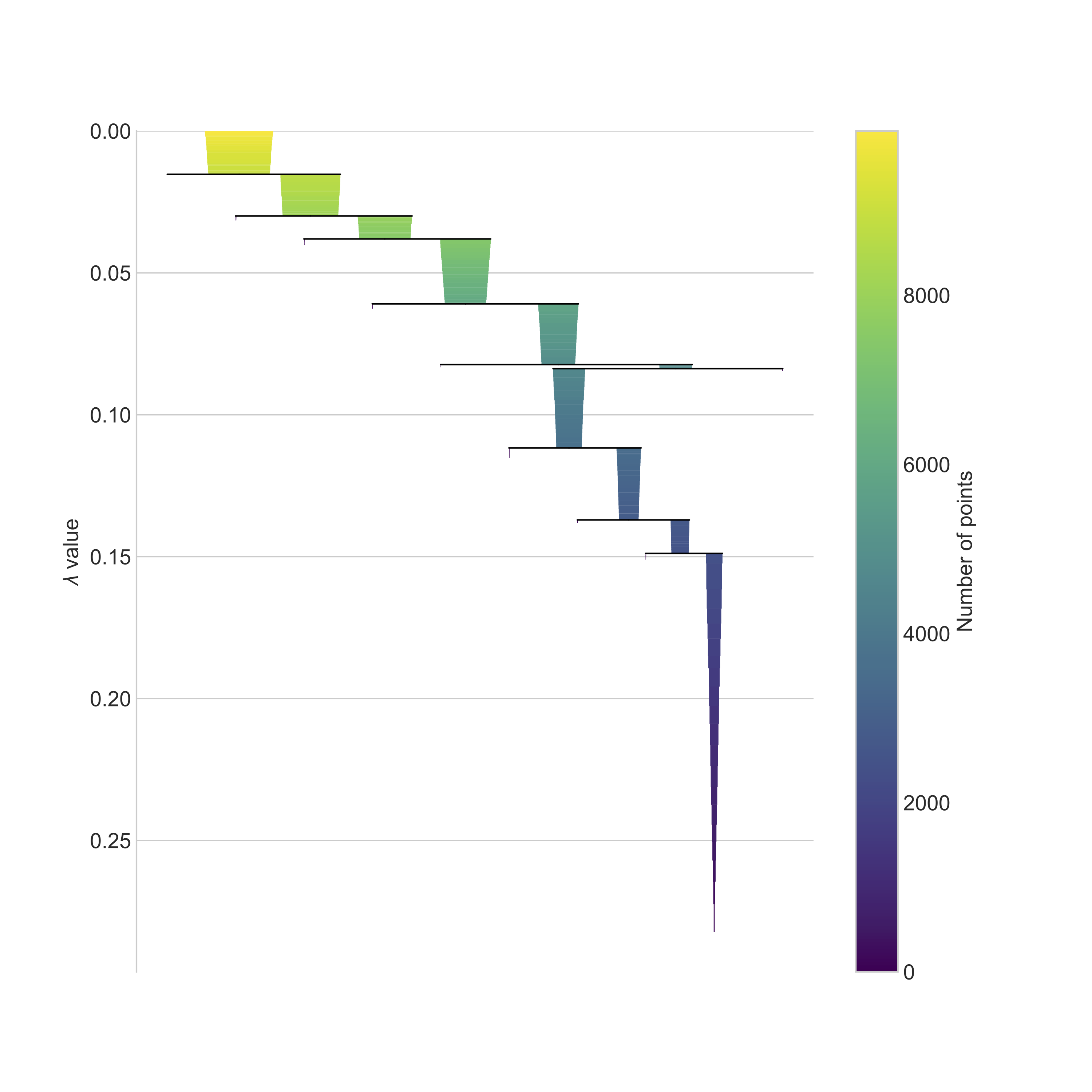}
  \caption{Condensed tree visualisation indicating the 10 persistent morphological clusters identified by HDBSCAN in a sample of the 10,000 brightest LoTSS-DR1 galaxies. Reading from top to bottom, the root contains all of the sample, and this branches off down the hierarchy; only clusters with a number of points exceeding a certain critical threshold (the minimum cluster size) persist, which is parameterised by the $\lambda$ value, which represents the inverse of the mutual reachability distance between points in the Haralick space (See Section 3.2 for further details).}
\end{figure}

\section{Data and Training}

\subsection{The LOFAR Two-metre Sky Survey}

To demonstrate the efficacy of Haralick features in representing radio galaxy morphology, we use the first data release from the LOw Frequency ARray (LOFAR) Two-metre Sky Survey (LoTSS), released in 2018 \citep[LoTSS-DR1]{proposal44}. LoTSS is a survey of the northern sky over 120--168\,MHz wide-area survey delivering continuum imaging at an angular resolution of $\theta=6''$. LoTSS-DR1 comprises just 2\% of the total survey area, with 424 square degrees of imaging covering the Hobby-Eberly Telescope Dark Energy Experiment (HETDEX) `Spring Field' (10$^{\rm h}$45$^{\rm m}$--15$^{\rm h}$30$^{\rm m}$ and 45$^\circ$--57$^\circ$). The median depth is 71$\mu$Jy\,beam$^{-1}$ at 144\,MHz, with a 90\% completeness for point sources at 0.45\,mJy. A total of 325,694 sources are catalogued across 58 mosaics covering the HETDEX field. We refer the reader to \cite{proposal44} for a thorough description of LoTSS and the DR1 data products.

Using the LoTSS-DR1 catalogue, we extract thumbnail cutouts around the position of the top 10,000 sources ranked by total flux. We adopt a window size of $64\times 64$ pixels ($96 \times 96$\,arcseconds) which we found to be appropriate for comfortably encompassing the majority of sources, with the exception of some of the very brightest and extended emission features. Each image is minmax normalised, put on an 8-bit grey scale and the Haralick features computed (note that we compute the features for all pixel offsets and average the result). Finally, each source is represented by a 13-element vector corresponding to the 13 Haralick features as described in Section 2. In the next section we describe an approach to cluster the feature vectors to provide groupings of galaxies with similar Haralick features.

\begin{figure*}
  \includegraphics[width=\textwidth]{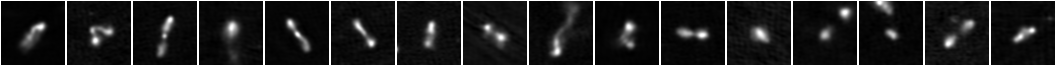}
  \includegraphics[width=\textwidth]{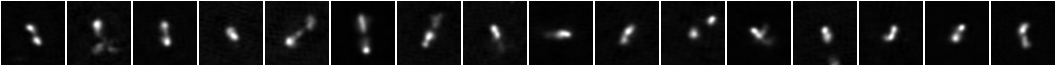}
  \includegraphics[width=\textwidth]{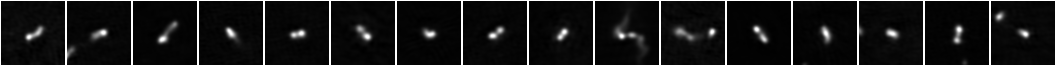}
  \includegraphics[width=\textwidth]{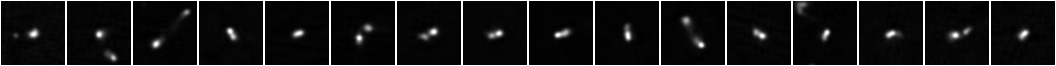}
  \includegraphics[width=\textwidth]{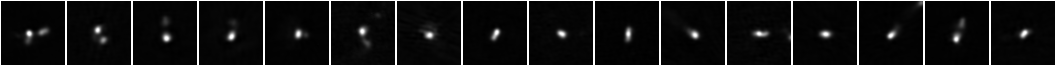}
  \includegraphics[width=\textwidth]{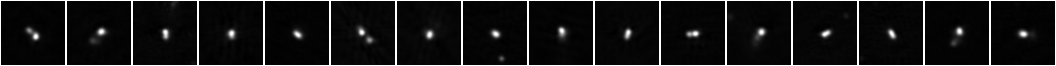}
  \includegraphics[width=\textwidth]{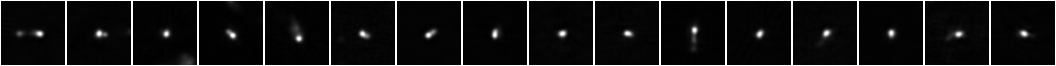}
  \includegraphics[width=\textwidth]{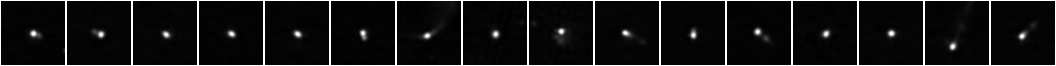}
  \includegraphics[width=\textwidth]{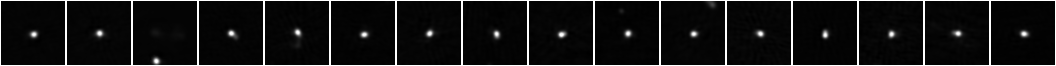}
  \includegraphics[width=\textwidth]{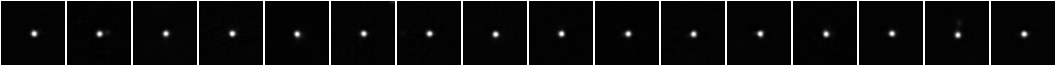}
  \caption{Thumbnails of LoTSS-DR1 sources sorted by similar Haralick features. Each row represents a cluster estimated by HDBSCAN (Section 3), and for clarity we show the top 16 examples (ranked by membership strength, with the first thumbnail in each row representing the most representative member). Each image subtends $96\times96$\,arcseconds and is linearly scaled between the 0.1-99.9th percentiles. The rows are sequential in terms of their extraction from the HDBSCAN hierarchical tree structure, and it is clear that the procedure essentially provides a morphological sorting ranging from radio galaxies with pronounced jets and lobes to point sources. Clustering of 10,000 objects took approximately half a second.}
\end{figure*}

 \subsection{Hierarchical Density-based Cluster Selection}
HDBSCAN is an unsupervised machine learning algorithm that is based on DBSCAN (Density-based
Spatial Clustering of Applications with Noise), first introduced by \cite{proposal42}. It is a density-based clustering algorithm that assumes clusters are characterised by `islands' of high density in the sea of the parameter space, such that clusters are regarded as data partitions that have a higher density than their surroundings. HDBSCAN \citep{proposal43} takes forward the DBSCAN concept by introducing a hierarchy to the clustering, with `persistent' clusters finally extracted from the hierarchical tree. Following \cite{proposal39}, HDBSCAN works as follows:
\begin{enumerate}
    \item The data is a set of vectors. In our case, each vector $x$ has 13-elements representing the Haralick features of each thumbnail image.
    \item Consider the `core' distance for a point $x$, $\text{core}_k(x)$, which defines the distance to the $k$th nearest neighbour, which is an efficient measure of density. Low values of $\text{core}_k(x)$ correspond to high density and vice versa.
    \item The `mutual reachability distance' between two points $a$ and $b$ is defined as $d_{\rm mut}(a,b)=\text{min}\{\text{core}_k(a),\text{core}_k(b), d(a,b)\}$. Here $d(a,b)$ is the distance between $a$ and $b$ according to some metric (e.g. Euclidean). The idea here is that the mutual reachability distance allows points in dense regions stay close together, and those in less dense regions to be pushed away. A mutual reachability graph can be used to represent the data set, with vertices as the data objects and weighted edges as connections. 
    
    \item The mutual reachability graph is used to construct the minimum spanning tree, and sorting its edges by the mutual reachability distance results in a hierarchical tree structure. The hierarchy of connected components is defined by sorting the edges of the tree by distance in reverse order, describing a dendogram.  This is the structure from which clusters will be identified. 
    
    \item We wish to extract `flat' clusters from the hierarchy, and this in principle achieved by slicing through the dendogram. Unfortunately, it is not clear at what point to make the cut, and if one does choose a single cut point (i.e.\ a given distance), that would effectively select clusters with the same density. HDBSCAN allows the extraction of clusters of variable density, effectively cutting the dendogram at different points.
    
    \item First the cluster tree is condensed into a simpler structure. Considering the single main trunk which contains all of the data points, the tree splits into branches. A condensed cluster hierarchy can be described by considering the number of points kept in each branch as it splits. Here we introduce the key parameter of minimum cluster size. If a given branch splits into two, with one branch containing fewer points than the minimum cluster size means, the larger branch `persists' and the smaller split branch `falls out' of the cluster. If a branch splits into two with both branches exceeding the minimum cluster size, both new branches persist, and so-on. 

    \item Clusters are extracted on the notion of persistence in the hierarchy. The parameter $\lambda=d_{\rm mut}^{-1}$ is defined, and each cluster has a $\lambda_{\rm birth}$ (the point at which the cluster split off) and $\lambda_{\rm death}$ (the point when the cluster split into other clusters). In each cluster, we have $\lambda_p$ describing when each point fell out of the cluster (or was split off into a new cluster), $\lambda_{\rm birth}\leq \lambda_p \leq \lambda_{\rm death}$. Cluster stability $S$ is defined as the sum of $\lambda_p - \lambda_{\rm birth}$ for all points in the cluster. To extract clusters the following procedure is followed. First, select all leaves as clusters. Then, working through the hierarchy, consider the stability of a parent cluster $S_p$ and its $n$ descendants $S^{0,1,2,...,n}_d$. If \smash{$S_p > \sum^n_{i=0} S^i_d$} we unselect all the descendants. If \smash{$S_p < \sum^n_{i=0} S^i_d$} then the cluster stability is set such that \smash{$S_p = \sum^n_{i=0} S^i_d$}. At the root node we have our set of selected clusters. Any point in the sample that does not fall into one of the selected clusters is defined as `noise'.
    
    \item The selected clusters can be used to label points. Furthermore, the definition of $\lambda_p$ within a cluster, when normalised between 0 and 1 provides a means of characterising a probability that a given point belongs to the cluster, or alternatively a measure of the strength of membership.
    \end{enumerate}
\noindent In this work we use the Python implementation of HDBSCAN\footnote{\url{https://hdbscan.readthedocs.io}}. As described above, there are some key parameters to set when applying the algorithm, and we experimented with their effect on the clustering results. We found that a simple Euclidean distance metric was effective and computationally efficient, and that other in-built distance metrics did not dramatically alter the results. The parameter `minimum cluster size' and `minimum number of samples' are the most critical. The former sets the minimum size of a cluster, as described in (v) above. We adopt a minimum cluster size of 64; in experimentation, lower sizes result in a larger number of clusters with very similar visual morphologies split over multiple labels and larger values result in too few clusters to make useful/meaningful classes. Our experimentation is based on visual inspection of the galaxies in the clusters returned for different minimum cluster sizes. The behaviour tends to be, as the minimum cluster size is reduced, some clusters are generated that contain a small (by definition) number of objects that are visually similar to some larger `main' cluster. To illustrate an example, if the minimum cluster size is reduced to 16, then we generate 40 clusters in total. The final two clusters -- by definition morphologically similar -- contain 196 and 3612 objects respectively, such that there is clearly a dominant cluster present. Since adjacent clusters are similar in the Haralick space, they are likely small `islands' broken off in the clustering space that now satisfy the more lenient minimum cluster size. As we describe in the following section, an adoption of `soft clustering' mitigates the need to define had cluster membership boundaries (in effect, every object has some finite probability of belonging to every other cluster).

In some sense, the total number of clusters required is a matter of taste or practical need. Indeed, one could argue that the desired number of labels \textendash i.e. the level of granularity in morphological distinction \textendash is use-case specific; for example, a limiting case might be the simple separation of point and extended sources, in which case just two clusters would suffice.The minimum number of samples controls how conservative the clustering is, manifested as the fraction of the data that is labelled as noise. This is generally set to the same value as the minimum cluster size. 

HDBSCAN is incredibly fast: for the 10,000 sample size considered in this demonstration, the clustering time was just 0.5\,seconds. Although performance is obviously hardware dependent, HDBSCAN offers high performance for the clustering of large samples, making the analysis of the very large radio galaxy {\it imaging} catalogues of the future (e.g.\ SKA and pathfinders) a tractable problem. In Figure~3 we show the scaling of clustering time with sample size for our 13-element features. 

\begin{figure}
    \centering
    \includegraphics[width=0.5\textwidth]{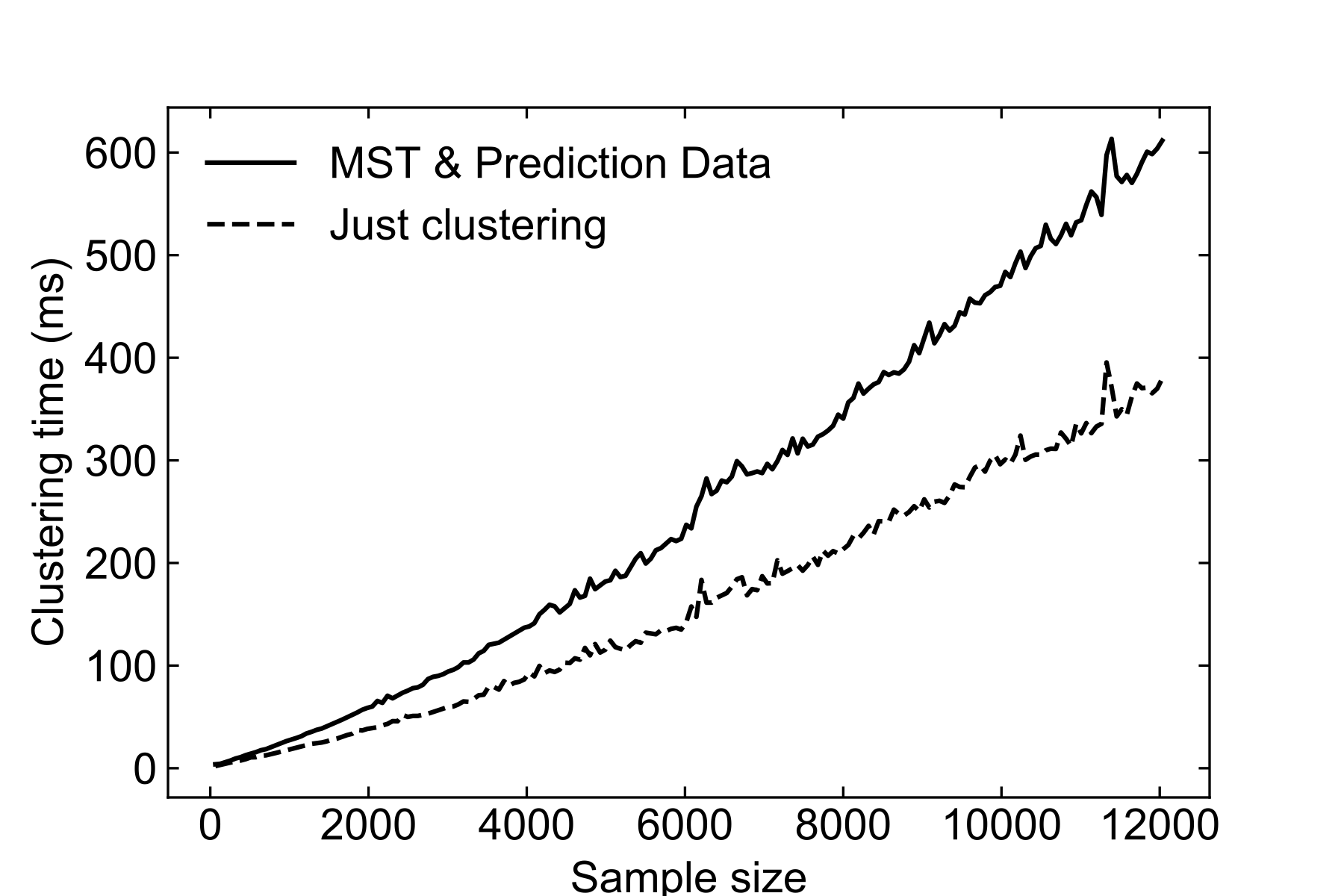}
    \caption{Clustering time as a function of sample size for the 13-element Haralick feature vectors. We compare the timing (in milliseconds) for clustering that includes the calculation of the minimum spanning tree (MST) and prediction data. The prediction data allow one to use the trained model to predict the labels of new, unseen feature vectors.}
\end{figure}

\begin{figure*}
    \centering
    \includegraphics[width=0.49\textwidth]{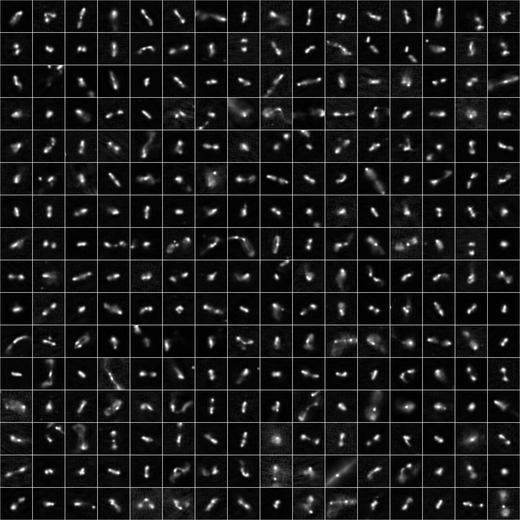}    \includegraphics[width=0.49\textwidth]{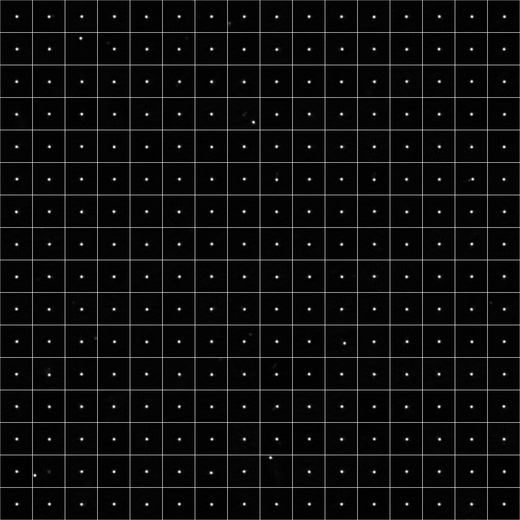}
    \caption{As Figure~2, but showing a larger sample of objects within cluster 1 (left) and cluster 10 (right). The images are organised row-wise in order of membership probability. The nature of the extraction of flat clusters from the condensed tree structure produced by HDBSCAN means that the sequence of cluster labels (in this case 1--10) forms a natural morphological sort, from highly extended and complex radio morphologies in cluster 1 to compact point sources in cluster 10.}
\end{figure*}

HDBSCAN has already been applied in astronomy. For transient discovery, \cite{proposal46} present an application in transient discovery. The authors evaluated 85,553 min-cadenced light curves from the Deeper, Wider, Faster program \citep[]{proposal51}. They were able to isolate anomalous sources for further analysis including the discovery of seven uncatalogued variables and two stellar flare events, including a rarely observed ultrafast flare. \cite{proposal45} used HDBSCAN to identify clusters of stars, galaxies, and quasars in a multidimensional space defined by photometry. Using a set of 50,000 spectroscopically labelled objects from the Sloan Digital Sky Survey \citep{proposal52} the authors found that careful attribute selection is a vital part of
accurate classification with hdbscan. They optimized the hyperparameters and input attributes of three
separate hdbscan runs, each to select a particular object class and, thus, treat the output of each separate run as a binary classifier. They then consolidated the output to give the final classification. They used F1 scores to measure the performance of their classifier and they obtained F1 scores of 98.9, 98.9, and 93.13 respectively. Finally, \cite{proposal48} presented the Milky Way Project second data release (DR2) and an updated data reduction pipeline. The authors aggregated about 3 million classifications from volunteers to produce the DR2 catalogue, which contained 2600 infrared `bubbles' and nearly 600 candidate bow shock-driving stars. HDBSCAN was used to create the bubble catalogue, identifying clusters within a set of one million classifications in less than 5 minutes on a standard desktop computer, again illustrating the speed performance of HDBSCAN when dealing with very large datasets.

\section{Results \& Discussion}

Since most of the morphological diversity is in the brightest radio sources (e.g.\ those with extended jets and lobes), in our demonstration we take the top 10,000 sources ranked by total flux in the LoTSS-DR1 catalogue. We run HDBSCAN with a minimum cluster size of 64 which results in 10 clusters containing 33\% of the sample. Figure\ 1 shows the condensed tree visualisation that shows the branching of the sample from the main root. Examples of the representative morphologies in each cluster are shown in Figure~2. Interestingly, the ordering of the cluster labels 1--10 form a progression: cluster 1 contains the most extended sources with prominent jets and lobes. As one examines successive clusters it is clear that the morphology transitions, encompassing more compact structures including close pairs of similar brightness (e.g.\ hotspots), pairs of sources dominated by one brighter component, and finally to isolated point sources. To emphasise this, in Figure~4 we show a larger sample of the top 256 most likely members of cluster 1 and cluster 10.

\begin{figure*}
  \includegraphics[width=\textwidth]{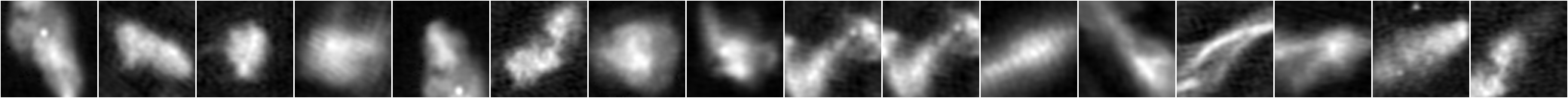}
  \includegraphics[width=\textwidth]{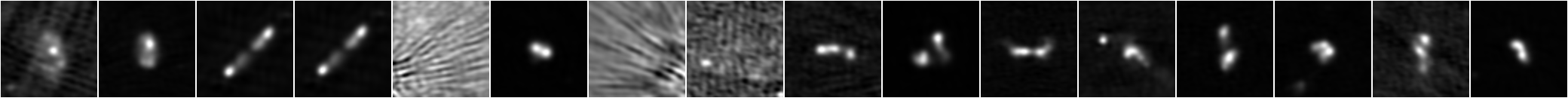}
  \includegraphics[width=\textwidth]{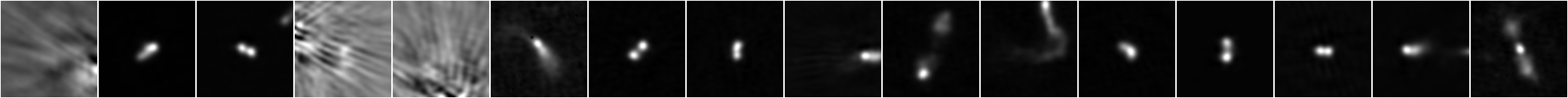}
  \includegraphics[width=\textwidth]{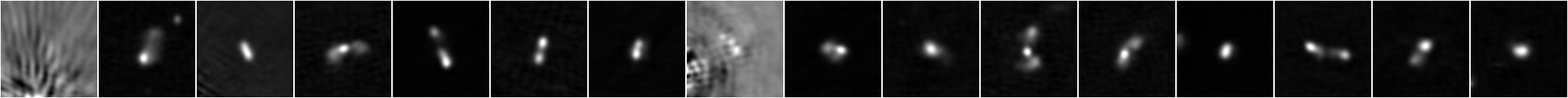}
  \includegraphics[width=\textwidth]{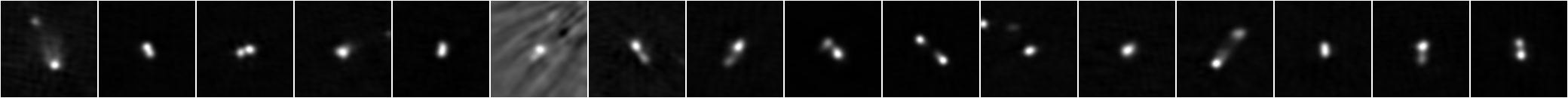}
  \includegraphics[width=\textwidth]{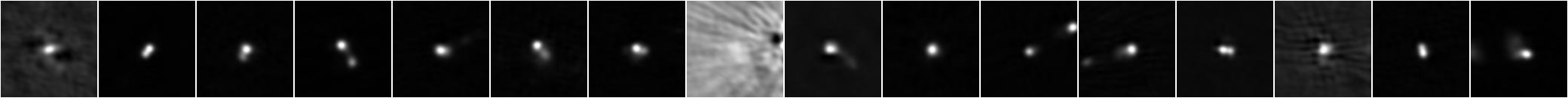}
  \includegraphics[width=\textwidth]{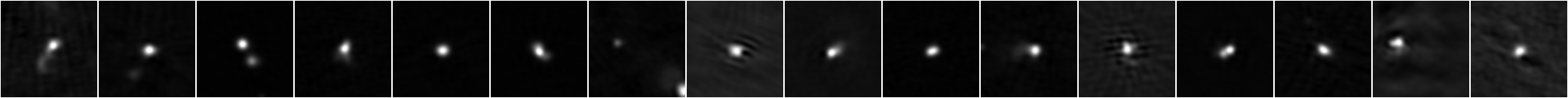}
  \includegraphics[width=\textwidth]{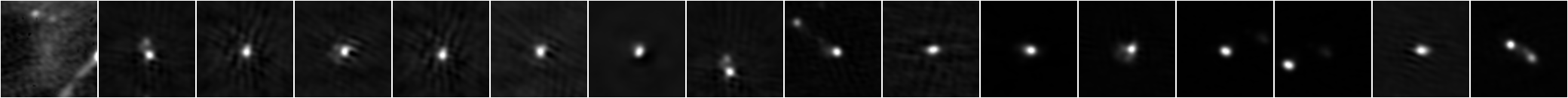}
  \includegraphics[width=\textwidth]{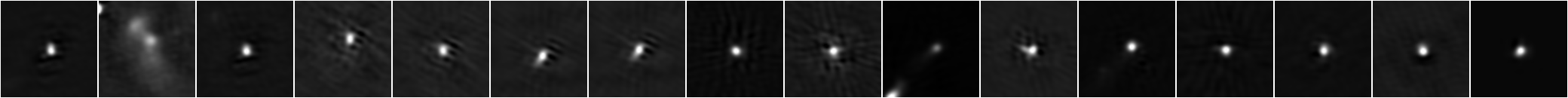}
  \includegraphics[width=\textwidth]{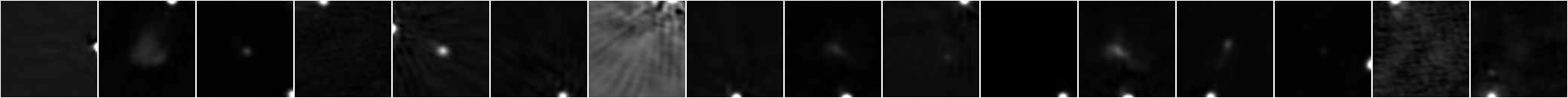}
  \caption{As Figure~2, but showing examples of galaxies assigned to clusters via soft clustering that HDBSCAN originally failed to classify (i.e.\ `noise'). In cluster 1 we can see that the sources are typically large and diffuse systems comparable to the size of the thumbnail. The general appearance of the sources in each row follows the trend seen in Figure~2, but it is clear that the sample contains `sources' contaminated by imaging artefacts and background features (e.g.\ ripples) that might help explain the reason they were originally classified as `noise' in the hard clustering.}
\end{figure*}

One drawback of HDBSCAN, at least in this application, is the large fraction of sources that are labelled as `noise' -- i.e.\ sources not assigned to any cluster. This is an outcome of the conservative nature of HDBSCAN and the `hard edge' nature of the clusters -- points close to the edge but just falling out of the selection are classified as noise. We could have achieved a lower noise fraction by reducing the minimum cluster size, such that a larger number of clusters would have persisted, however this has the drawback that the cluster labels become less useful. An alternative approach, which we take here, is to use the concept of `soft' clustering to assign {\it every} source to its most likely cluster. Soft clustering avoids the hard boundaries imposed by traditional clustering algorithms, and instead takes a probabalistic view: each point has a probability of belonging to a given cluster. In HDBSCAN, soft clustering is achieved via a modification of the outlier score, which is based on the Global-Local Outlier Score from Hierarchies (GLOSH) algorithm (Campello et al.\ 2015), and combines this with a measure of distance from a given cluster to produce an estimate of the probability that any given point belongs to any of the fixed clusters extracted from the condensed tree. We can then simply assign cluster labels for every point by taking the most likely cluster it belongs to, and extract samples of galaxies for each label using probability thresholds. This fuzzy clustering approach is attractive because it recognises the fact that real galaxies may have shared characteristics (e.g.\ jet plus point source versus jet without point source). In effect, it is the distribution of membership probabilities for a given source that describes the source morphology. Parameterising morphology this way allows more complex morphological-based selections to be made that combine the full `spectrum' of morphological types.

In Figure~5 we duplicate Figure~2, but this time show examples of sources originally classified as noise by HDBSCAN. Soft clustering allows us to assign the most likely cluster membership, but this visualisation helps us understand the nature of these `noisy' galaxies. In cluster 1 (top row) we can see that the sources are all rather large and diffuse radio galaxies; real sources but rare in the catalogue and therefore likely fall foul of the minimum cluster size to be designated an independent cluster. In the sequence of rows corresponding each cluster label we can see the same trend of `extended/lobed' systems through to `compact/point-like' systems as in Figure~2, indicating that the soft clustering assignment still respects the original classification scheme. However, thumbnails often exhibit contamination from artifacts (e.g.\ from the interferometric imaging, such as ripples, striations and halos due to nearby bright sources), or the presence of low signal-to-noise features such as diffuse emission. In these cases we would expect the Haralick features to encode the image textures associated with these properties. The soft clustering assignment, along with the book-keeping regarding the sources' original classification allows us to easily identify such outliers.

How do the different clusters compare to the basic observational properties of the members? We consider cluster members in each of the 10 clusters with high $>$90\% membership probabilities and compare the distributions of total flux, major axis and ratio of major to minor axis as described by \cite{proposal44}.\footnote{ The fraction of sources with $>$90\% membership probability across all clusters is 10\%. At $>$50\% membership probability the fraction is 40--50\%.} Figure~6 compares the distributions. The distribution of flux within each cluster is remarkably consistent, although there is some variation in the bright tails of each cluster's distribution. As expected, there is a stronger correlation with the shape measurements (although notably broad and overlapping distributions in each case). For example, there is a progression with the numerical cluster label (which, recall, is dependent on the point in the tree where the cluster was extracted, see Figure~1). For example, the most extended sources are in cluster 1, with the most compact in the final cluster 10. It is clear that it would be challenging to reproduce the same morphological groupings simply by using the basic catalogue measurements of flux density, major and minor axis. Finally, we note that, since clusters are extracted in a sequential fashion from the tree, and that the characteristics of sources in cluster 10 are `far away' from those in cluster 1 (in Haralick space) means that HDBSCAN has produced a meaningful sorting of radio galaxies by morphology.    

The Haralick features are trivial to compute direct from imagery via the GLCM, and -- as demonstrated here -- essentially provide a compact description of source morphology. The Haralick feature values could be trivially included in source catalogues at relatively little memory expense. Care must be taken however: since the features are calculated from pixel values, they cannot be easily cross-compared between surveys (for example), since they also encode information about the noise properties of the image. That effect is not examined in detail here since we have focused on very high signal-to-noise sources for the present demonstration.

\begin{figure}
\includegraphics[width=0.5\textwidth]{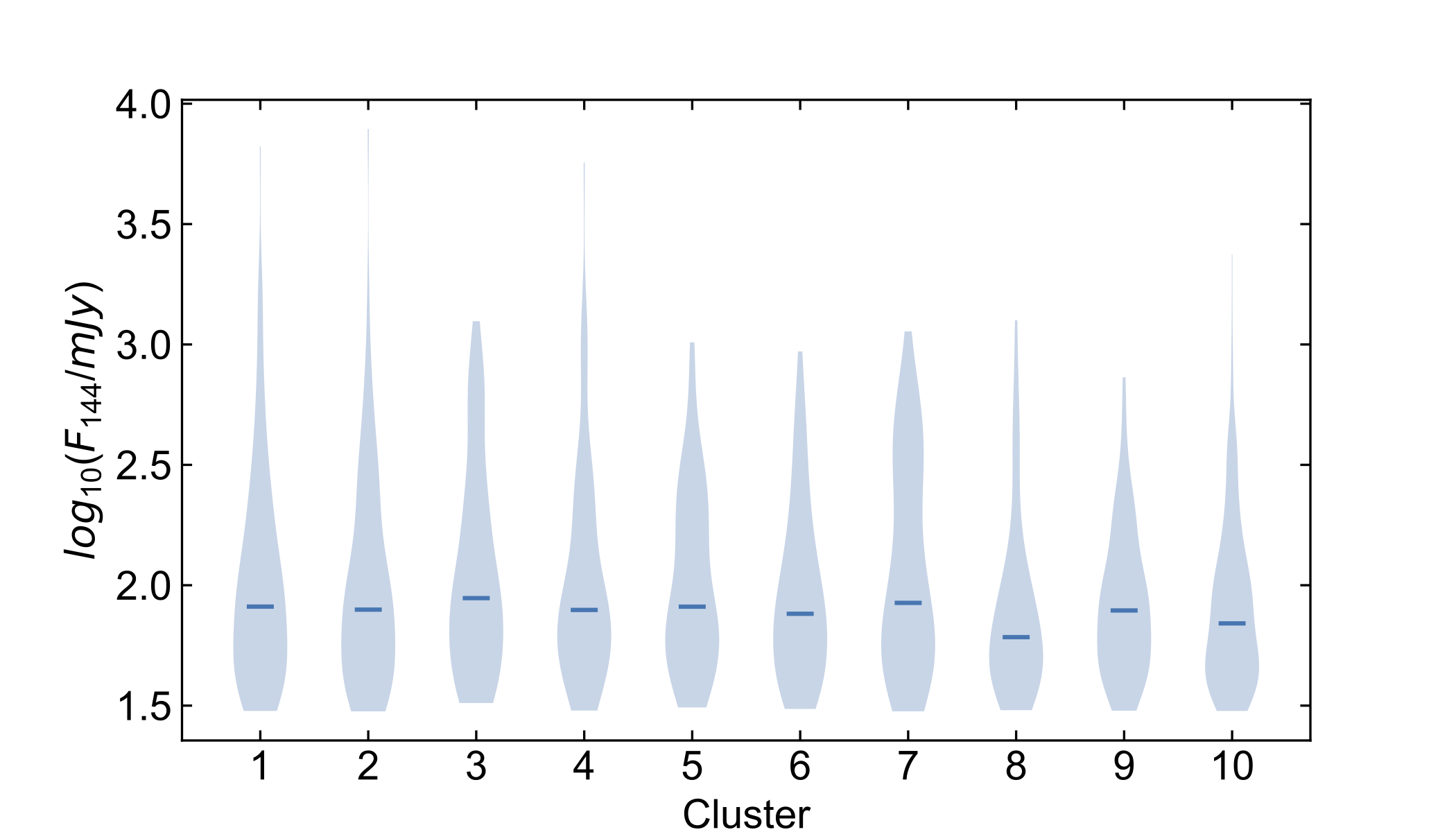}
\includegraphics[width=0.5\textwidth]{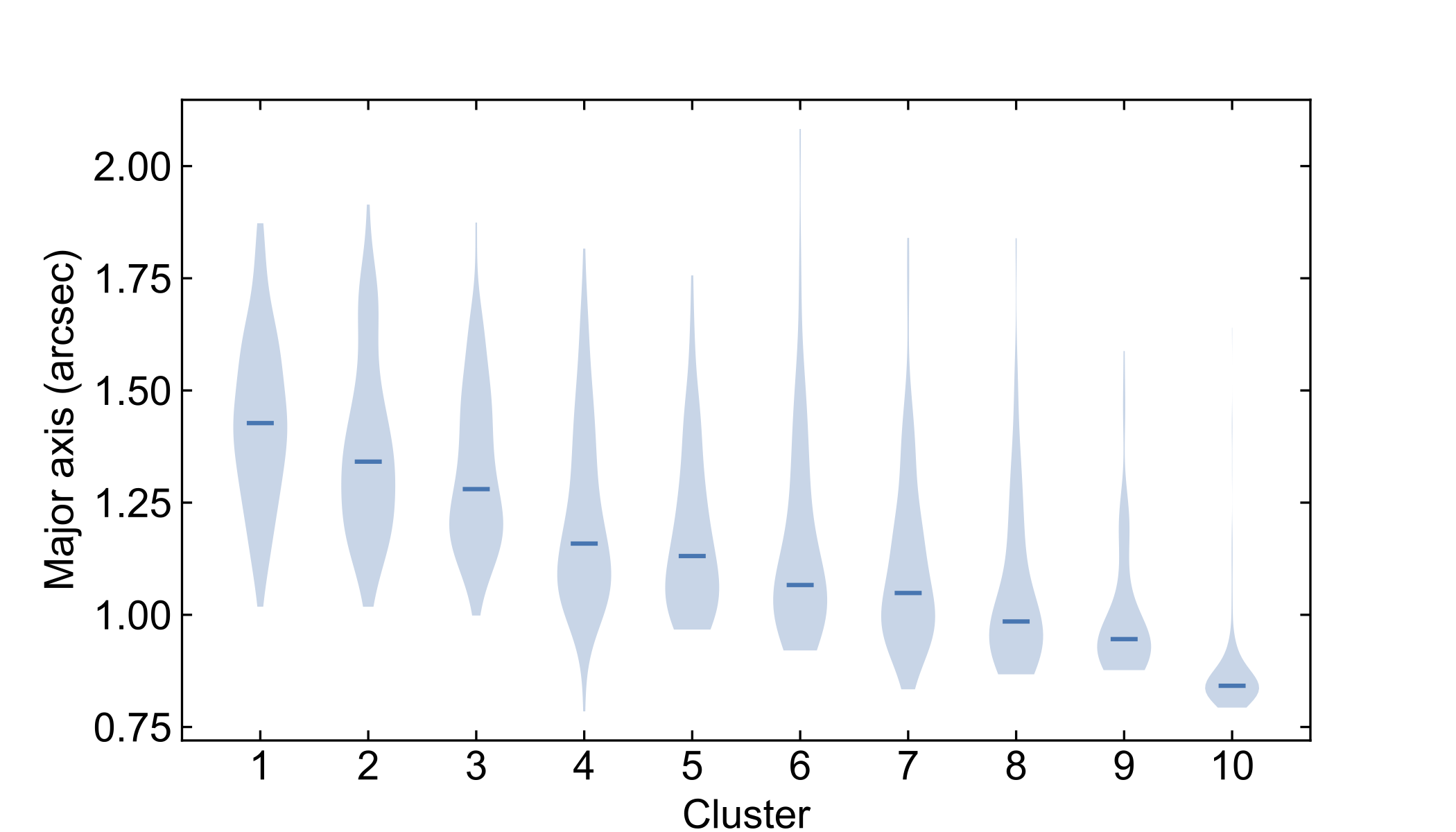}
\includegraphics[width=0.5\textwidth]{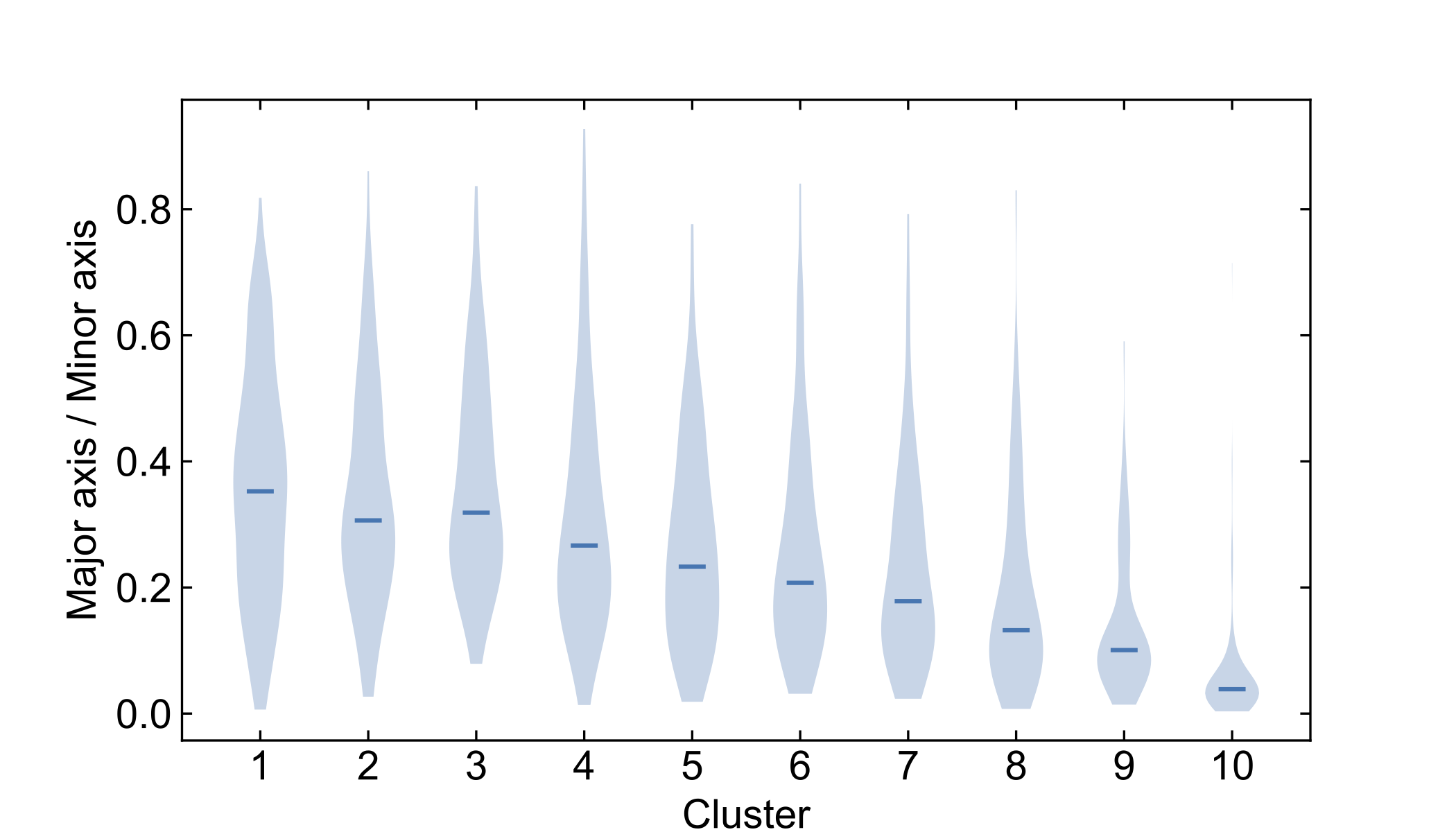}
\caption{A comparison of the distribution of (top) total 144\,MHz flux, (middle) major axis and (bottom) ratio of major to minor axis for cluster members with a probability of membership exceeding 90\%. The hard cut offs in the flux distributions is a result of the original flux-based selection. Horizontal lines indicate the median values. }
\end{figure}

 \section{Conclusions}
 
With a focus on radio galaxies, we have shown how Haralick features can be used as simple, compact and non-parametric descriptors of source morphology. Haralick features are an established tool in computer vision, and simply encapsulate local greyscale image texture with 13 numbers evaluated from the Grey Level Co-occurance Matrix. Since morphological variation is in essence due to differences in the spatial distribution of pixel intensities in imaging, our hypothesis was that Haralick features could be used to represent morphology.

Using sources from the LOFAR Two-metre Sky Survey \cite{proposal44}, we calculate Haralick features within 96$''$ apertures for the 10,000 brightest catalogued sources, spanning extended systems with prominent lobes and jets, to compact point sources. We find distinct morphological clusters within the `Haralick space' using the fast HDBSCAN clustering algorithm, which distinguishes ten main morphological classes. Rather than fixing cluster assignment, we adopt soft clustering which, rather than giving points a specific cluster label, provides a probability of cluster membership. This allows for far more flexibility in morphological selection, given that often galaxies will share morphological characteristics across multiple labels. As clusters are extracted from a hierarchical structure, and the distance between clusters in the hierarchy depends on morphology (as encoded by Haralick features) the sequence of cluster labels provided by HDBSCAN are sorted in a meaningful way. In our example, cluster 1 contains the most extended sources with prominent lobes and jets, and the final cluster 10 contains isolated and compact sources. The probabalistic nature of the cluster assignments also allow one to identify morphological outliers, making the detection of unusual galaxies (or artifacts) trivial. HDBSCAN offers another advantage: it is an incredibly fast clustering algorithm. In our demonstration, the clustering and classification of 10,000 sources took just half a second. Here we focus on radio morphologies, but of course Haralick features could be calculated for any set of imaging data (e.g.\ optical), and can be extended to more than two dimensions, allowing one to encode spectral/colour information as part of the classification.    

The exploration and curation of the giant imaging datasets of the future (e.g.\ SKA in the radio, LSST in the optical) requires efficient methodologies, both in terms of computational expense and storage considerations. We argue that Haralick features are a valuable addition to non-parametric methods of morphological classification of galaxies, with hierarchical soft clustering of the Haralick features of a given sample providing a convenient means of sorting and classifying galaxies in the morphological parameter space.

\section*{Acknowledgements}
KN is supported by the Development in Africa with Radio Astronomy (DARA-BIG DATA) scheme which is a Newton Fund project delivered through the UK Science and Technology Facilities Council (STFC) (ST/R001898/1). JEG acknowledges support from the Royal Society (URF/R/180014). 

\section*{Data Availability}
The data and code used to produce the results presented in this paper can be found on Github at \url{https://github.com/KushathaNtwaetsile/HaralickFeatures}.




\bibliographystyle{mnras}
\bibliography{reference}

\bsp	
\label{lastpage}
\end{document}